# Prioritizing App Reviews for Developer Responses on Google Play

Mohsen Jafari*
*Department of Computer Science*
*Bowling Green State University*
Bowling Green, USA
mjafari@bgsu.edu

Forough Majidi*
*Department of Comp. and Soft. Eng.*
*Polytechnique Montréal*
Montréal, Canada
forough.majidi@polymtl.ca

Abbas Heydarnoori
*Department of Computer Science*
*Bowling Green State University*
Bowling Green, USA
aheydar@bgsu.edu

## Abstract

*The number of applications in Google Play has increased dramatically in recent years. On Google Play, users can write detailed reviews and rate apps, with these ratings significantly influencing app success and download numbers. Reviews often include notable information like feature requests, which are valuable for software maintenance. Users can update their reviews and ratings anytime. Studies indicate that apps with ratings below three stars are typically avoided by potential users. Since 2013, Google Play has allowed developers to respond to user reviews, helping resolve issues and potentially boosting overall ratings and download rates. However, responding to reviews is time-consuming, and only 13% to 18% of developers engage in this practice. To address this challenge, we propose a method to prioritize reviews based on response priority. We collected and preprocessed review data, extracted both textual and semantic features, and assessed their impact on the importance of responses. We labelled reviews as requiring a response or not and trained four different machine learning models to prioritize them. We evaluated the models' performance using metrics such as F1-Score, Accuracy, Precision, and Recall. Our findings indicate that the XGBoost model is the most effective for prioritizing reviews needing a response.*

***Index terms*—** Prioritizing App Reviews, Mobile Applications, Machine Learning, Sentiment Analysis, App Stores.

## 1 Introduction

In recent years, the number of mobile apps in app stores, including Google Play, has increased dramatically. In the Google Play store, users can express their opinions and rate apps [1, 19]. These reviews include reports of problems and suggestions for app improvement. Also, app ratings indicate users' satisfaction and affect the number of downloads and the success of an app [16]. Studies show that the app rating is a principal factor for users. Apps with a rating of less than three stars are not downloaded by 77% of users [17]. To respond to user needs better, Google Play introduced a feedback mechanism for developers in 2013. To avoid receiving low ratings, developers must respond to user reviews. However, responding to reviews is time-consuming and costly, and less than 1% of reviews are responded to according [17]. Based on our knowledge, no research has been done on prioritizing user reviews to respond. Therefore, in this research, we have addressed the existing gap in prioritizing user reviews to respond. Two criteria are defined to measure the importance of reviews. The first criterion is whether or not to respond to reviews, and the second criterion is used to determine reviews that need to be responded to urgently (high-priority reviews).

The approaches presented in this paper include the following main steps. First of all, user reviews and developer responses are extracted from the Google Play store and are preprocessed. Secondly, textual and semantic features are extracted for each review and are measured the impact of each feature on the importance of the response to the review. Then we labeled our data based on response or not necessary response. After training four machine learning algorithms, their performances are evaluated. Finally, the evaluations of several different apps with different properties indicate that the XGBoost algorithm shows the best performance.

---

*These authors contributed equally to this work.

The proposed approaches in this paper have the following main benefits for developers: (i) the volume of reviews that the developer has to read and respond to is reduced; (ii) Help developers decide which reviews to respond to; and (iii) developers are helped to determine the high priority reviews to respond.

The main contributions of this work include: (i) studying the overall relationship between textual and semantic features of reviews with developer responses, and identifying and introducing the most influential features in the importance of reviews. This aims to perform more accurate and comprehensive predictions of developer behaviour. Additionally, (ii) comparing the performance of four machine learning algorithms for solving the problem of prioritizing user reviews for responses, and selecting the best algorithm.

This paper is organized as follows. Section 2 presents the proposed approaches. Section 3 provides the evaluations of the proposed approach. Section 4 discusses the strengths and weaknesses of the proposed approach. Section 5 provides an overview of related work. Finally, section 6 concludes the paper and provides future research directions.

## 2 Proposed Solution

### 2.1 Approach Overview

We aim to develop a system for prioritizing user reviews to respond. The proposed solution involves several steps outlined in Figure 1. Initially, data is collected by selecting popular apps from Google Play and those referenced in previous research. Web crawling is utilized to extract initial information, which is then merged with existing datasets. In the preprocessing stage, reviews undergo various treatments such as language filtering, punctuation removal, and stemming. Textual and semantic features are then extracted in the third step to describe the reviews. Reviews are labeled based on whether a developer response is provided or not. Four machine learning algorithms are employed, including Decision Tree, Random Forest, Support Vector Machine (SVM), and XGBoost to model developer behavior. The impact of features on the output is assessed, and models are trained and evaluated for the best performance and result.

### 2.2 Data Collection

We aim to select datasets from a variety of app categories, such as photography, news, magazines, maps and navigation. In the first step, several popular apps on Google Play Store are identified. Then, all the information of these apps, including reviews and responses, is extracted by a web crawler. The collected data is merged with the dataset from the work presented in [7]. We gathered 431512 reviews that were responded to or not responded to. The dataset consists of the following 7 columns: App Name, Review, Review Score, Review Submission Date, Total Useful Score of the Review, Response, and Response Submission Date. The structure of the dataset is illustrated in Table 1

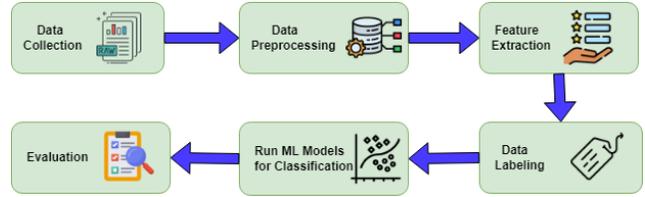

Figure 1: Approach overview

Table 1: The structure of the dataset

| No. | Column Name | Description |
| --- | --- | --- |
| 1 | App Name | The name of the unit that each app has |
| 2 | Review | User feedback about the app |
| 3 | Review Rating | The rating that the user gives to the app from 1 to 5 |
| 4 | Review Submission Time | The date the user submitted their review |
| 5 | Total Helpful Votes | The number of people who found the review helpful |
| 6 | Response | The response that the developer registers for a review |
| 7 | Response Submission Time | The date the developer responded to the review |

### 2.3 Data Preprocessing

The reviews are in natural language and are posted by individuals with different languages. Also, those who post these reviews may have spelling mistakes due to time constraints. Therefore, it is necessary to preprocess the input data. It should be noted that in this research, preprocessing has been done at four levels: character, word, sentence, and review. At the character level, punctuation marks are removed. Moving to the word level, stop words are eliminated, and letters become lowercase. Additionally, spell-checking is conducted, and roots are identified through morphological analysis, followed by general root finding. Single-letter words are then removed. At the sentence level, sentences are tokenized by breaking them down into individual tokens. Finally, at the review level, non-English reviews are removed, and reviews are segmented into sentences for further analysis. Most of the functions we use for preprocessing are from the NLTK library in Python.

### 2.4 Feature Selection and Extraction

In the field of machine learning, features are seen as measurable characteristics of an event. Selecting independent and informative features is a fundamental step in using machine learning algorithms. After selecting suitable features to describe the event, they are prepared as feature vectors and fed into machine learning algorithms. It is necessary to transform the reviews into a format understandable by the computer. For this purpose, reviews need to be described using various features. These features are divided into two categories: textual and semantic which are introduced in Table 2.

Table 2: Textual and Semantic Features of Reviews

| Feature No. | Feature Type | Feature Name |
|---|---|---|
| 1 | Textual | Clarity level of the review |
| 2 | Textual | Length of the review |
| 3 | Textual | Complexity level of the review |
| 4 | Textual | Number of nouns in the review |
| 5 | Textual | Number of verbs in the review |
| 6 | Textual | Number of adjectives in the review |
| 7 | Textual | Number of adverbs in the review |
| 8 | Semantic | Sentiment of the review |
| 9 | Semantic | Neutrality level of the review |
| 10 | Semantic | Polarity level of the review |
| 11 | Semantic | Rating of the review |
| 12 | Semantic | Total usefulness score of the review |
| 13 | Semantic | Informative level of the review |
| 14 to 18 | Semantic | Purpose of the review |
| 19 | Semantic | Title of the review |
| 20 to 24 | Semantic | Number of commitment-expressing words |
| 25 to 34 | Semantic | Review inclinations |

### 2.4.1 Textual Features

Textual features provide us with information about the text of the review, but they do not provide any information about the content of the review. The *readability of review* will cause the developers to choose reviews that are more readable due to time constraints [12]. If the *review length* is long, the more likely it contains important information [10]. For example, the two review samples "bad" and "bad user interface, bad graphical design, the worst GUI." are different in terms of their level of difficulty, which may affect how they are categorized by developers. Therefore, in this study, *review complexity* is chosen as a feature to describe the review. In [12], the total number of nouns and verbs is introduced as a feature to describe reviews on the social web. [3] also states that nouns and verbs convey the majority of the meaning of a text. Therefore, the *number of nouns in review* in each review is chosen as a textual feature to describe it. Also, there are considered textual features for the number of verbs, adjectives, and adverbs.

### 2.4.2 Semantic Features

Semantic features in this study are descriptors related to the content of reviews rather than the text itself. The features extracted include the sentiment of the review, neutrality, polarity, review rate, total helpfulness score, informativeness, review purpose, review title, number of words expressing commitment, and review tendencies. *Sentiment analysis* is crucial as users may give positive reviews with low ratings due to misinterpretation of how to rate an app correctly [7]. *Neutrality* and *polarity* levels are introduced as features characterizing reviews [12]. The *review rate* on Google Play, ranging from 1 to 5 stars, reflects user satisfaction and influences developer response likelihood [10]. The *total helpfulness score* indicates the perceived usefulness of a review. *Informativeness*, determined by TF-IDF, gauges the richness of opinions. *Review purpose* categorization includes informative, feature request, problem discovery, etc [26]. *Review titles* are considered significant descriptors [24]. The *count of commitment words* like "must" indicates persuasive elements [4]. *Review tendencies* encompass categories like anger, sadness, and positive/negative feelings, each quantified by word count [20]. The article [24] highlights that user reviews and developer responses have distinct sections with varying *Topics* and purposes, influencing the responses. On Google Play, users review diverse topics such as UI, security, and pricing. This research used the LDA (Latent Dirichlet Allocation) model to extract and categorize review topics. The LDA model, trained on different numbers of topics, found 10 to be optimal. The scikit-learn library in Python was utilized to implement the LDA algorithm and calculate topic assignment probabilities. These features collectively contribute to a comprehensive understanding of review content and aid in prioritizing them for response.

## 2.5 Data Labeling

This section explains the labelling process of the collected data. We use supervised learning methods, which require data to be labelled. Our main goal is to prioritize user reviews to respond. There are two approaches to do the labelling:

1. In the first approach, reviews that have been responded to by developers are given more importance and receive a label of 1, while reviews that have not been responded to receive a label of 0.

2. In the second approach, reviews that have been responded to more quickly (three days or less) by developers are given more importance and receive a label of 1, while reviews that have been responded to in a longer time frame receive a label of 0.

## 2.6 Identifying Influential Features

In this study, machine learning algorithms convert training data into feature vectors for model input [23]. The Pearson correlation coefficient is used to analyze the correlation between features and the output value, as well as the correlation among features themselves. If multiple features are highly correlated, redundant ones are ignored. The best model is then selected based on its performance in prioritizing reviews.

## 2.7 Machine Learning Models

In this study, four classification algorithms, namely Decision Tree, Random Forest, Support Vector Machine, and XGBoost have been used to learn developer's behavior and predict reviews that are of greater importance. Finally, a comparison has been made on the performance of these four models, and the best model has been selected. It is worth mentioning that the reason for selecting these four algorithms is their usage in other classification problems [2]. Additionally, these models represent the state of the art for this kind of dataset, with each employing a completely different algorithm. They are lightweight, resource-efficient, and have acceptable runtime execution. Once the classification algorithms were determined, we proceeded to train the models and evaluate their performance.

## 2.8 Prioritizing Reviews

This task is carried out from two perspectives. Firstly, it is examined whether the target review is worth responding to or not which we call Approach #1. For this purpose, after preprocessing and extracting textual and semantic features, machine learning methods are used to learn developers' past behaviour. The best model is selected for this perspective. Secondly, the importance of the review to the developer is indicated, and the decision is made on whether to respond quickly or not which is called Approach #2. After training the machine learning models, the best model for prioritizing reviews from this perspective is selected. This model assists the developer in deciding whether to respond to the review promptly or defer the response to another time.

### 2.8.1 Approach #1: Prioritizing Reviews as Requiring a Response or Not

In this approach, it is analyzed whether or not user reviews should be responded to. A dataset of 123,130 labeled reviews has been analyzed. After preprocessing and labeling the data, the textual and semantic features have been extracted from the reviews. Table 3 shows the correlation of each feature with the output variable in Approach #1.

After examining the correlation of each feature with the output variable and extracting the features that have the greatest impact on the output, we will train our models with selected features. Table 3 illustrates the features that we will use for Approach #1, highlighted in bold such as score, length, or the readability of the review.

Table 3: Abbreviated Name, Full Name and Correlation of Extracted Features with Output for Prioritizing Reviews for Both Approaches

| Abbr | Feature Name | Approach #1 | Approach #2 |
|---|---|---|---|
| F1 | Total Useful Review Score | 0.031 | - |
| F2 | Review Score | **-0.51** | **0.34** |
| F3 | Review Length | **0.22** | 0.12 |
| F4 | Review Readability | **-0.16** | **0.2** |
| F5 | Review Complexity | 0.21 | 0.13 |
| F6 | Neutrality | **-0.12** | **0.56** |
| F7 | Polarity of Review | **0.23** | **0.42** |
| F8 | Number of Nouns | 0.19 | 0.15 |
| F9 | Number of Verbs | 0.026 | 0.33 |
| F10 | Number of Angry Words | 0.025 | 0.36 |
| F11 | Number of Sad Words | 0.007 | 0.36 |
| F12 | Number of Anxious Words | 0.023 | **0.36** |
| F13 | Number of Negative Words | 0.045 | 0.36 |
| F14 | Number of Positive Words | 0.054 | **0.39** |
| F15 | No. of Words Commitment | -0.26 | 0.36 |
| F16 | Review Sentiment | 0.022 | 0.27 |
| F17 | Topic 0 | **0.12** | 0.4 |
| F18 | Topic 1 | 0.099 | 0.41 |
| F19 | Topic 2 | -0.021 | 0.41 |
| F20 | Topic 3 | -0.023 | 0.4 |
| F21 | Topic 4 | 0.028 | **0.4** |
| F22 | Topic 5 | **-0.17** | 0.41 |
| F23 | Topic 6 | 0.087 | 0.42 |
| F24 | Topic 7 | -0.015 | 0.4 |
| F25 | Topic 8 | -0.063 | 0.41 |
| F26 | Topic 9 | **0.12** | 0.41 |
| F27 | Number of Adverbs | 0.18 | - |
| F28 | Number of Adjectives | 0.26 | - |
| F29 | Review Informativeness | 0.033 | - |
| F30 | Feature Request Category | 0.092 | **0.36** |
| F31 | Problem Detection Category | 0.014 | **0.36** |
| F32 | Information Request Category | **0.13** | **0.36** |
| F33 | Informer Category | **-0.17** | **0.38** |
| F34 | Other Review Categories | **1** | **0.83** |

Additionally, the features that are dependent on each other were also identified. It should be noted that, among the dependent features, the feature that has the highest correlation with the output variable is kept and the other features are removed from training the models. In this approach the features F3, F5, and F8 have strong correlations over 0.9, so we just consider F3 for our model.

### 2.8.2 Approach #2: Prioritizing Reviews as High or Low Priority for Response

In this approach, we determine whether or not a review needs to be responded to urgently (i.e., in less than three days) or not. For this purpose, data labelling is conducted differently from the previous section. Initially, a dataset of 308,382 reviews is earmarked for implementation and evaluation. Subsequently, reviews updated after receiving a response are removed due to a lack of initial post dates, leaving 284,062 reviews. The average response time to reviews of the dataset is calculated at 3.76 days, with reviews receiving responses within 0-3 days labelled as 1 and those after 4 days as 0. Data is then preprocessed before extracting textual and semantic features for each review. Notably, four features (useful votes, informative votes, adverbs, and adjectives) are excluded. Pearson correlation coefficient is employed to calculate the correlation between features and output, as well as between features themselves. The correlation of each feature with the output variable is depicted in Table 3. Subsequently, our training models are evaluated using 5-fold cross-validation, ensuring that the data is split into five subsets, with each subset being used as a test set once while the remaining four subsets are used for training. No feature exhibits a correlation of less than 0.1 with the output variable. However, certain features show high interdependence, with features F8 to F11, F13, and F15 to F28 having correlations exceeding 0.9. Among them, only feature F21, with the highest correlation with the output variable, is retained, while the others are discarded.

Table 4: Results of the performance of the four selected models for Approach #1

| Model | Accuracy | F1 Score | Recall | Precision |
|---|---|---|---|---|
| Decision Tree | 0.71 | 0.71 | 0.71 | 0.71 |
| Random Forest | 0.63 | 0.49 | 0.63 | 0.73 |
| SVM | 0.75 | 0.74 | 0.75 | 0.75 |
| XGBoost | **0.77** | **0.78** | **0.77** | **0.77** |

Table 5: Results of the performance of the four selected models for Approach #2

| Model | Accuracy | F1 Score | Recall | Precision |
|---|---|---|---|---|
| Decision Tree | 0.86 | 0.85 | 0.86 | 0.85 |
| Random Forest | 0.26 | 0.32 | 0.26 | 0.86 |
| SVM | 0.91 | 0.87 | 0.91 | 0.83 |
| XGBoost | **0.91** | **0.87** | **0.91** | **0.86** |

## 3 Evaluations

This section represents the results obtained from training four machine learning models for two approaches which can be seen in Tables 4 and 5. The performance metrics we used in this research are Accuracy, F1-Score, Recall, and Precision.

### 3.1 Evaluations of Approach #1

In this section, the performance of the models trained in Section 2.8.1 will be evaluated.

After determining the features that had a correlation of more than 0.1 with the output variable and were also not dependent on any other feature such as review score, review length, and review readability, the selected models were trained. The first part of Table 4 shows the results obtained from the performance of each of the models when considering only the selected features.

Table 4 shows that the XGBoost model has a higher F1-Score than other models for this approach with a value of 0.77. The other performance metrics such as Accuracy with a value of 0.78, Recall with a value of 0.78, and Precision with a value of 0.77 have higher values as well. According to the results obtained in this section, it can be concluded that the proposed approach in this paper for prioritizing user reviews to respond from the aspect of whether or not to respond to a particular review works well and has achieved very good and acceptable results.

#### 3.1.1 Case Study Evaluations

In this section, the best model obtained was tested on 10 popular applications. Firstly, ten popular applications from ten different categories were selected, and then 2000 comments were randomly selected for each application. After selecting the applications, the best model obtained was evaluated on these ten applications. The mean of F1-Score for theses 10 applications is 0.77 that shows the acceptable performance of the selected model. Also, it can be concluded that the presented model can be used for different types of applications from different categories.

### 3.2 Evaluations of Approach #2

In this section, the performance of the models trained in Section 2.8.2, which aim to prioritize reviews from the perspective of the importance of responding to them more quickly (3 days or less than 3 days), will be evaluated.

After determining the features that were not dependent on any other feature like review readability, neutrality, and polarity of review, the selected models were trained. The second part of Table 5 shows the results of the performance of each model.

Table 5 reveals that the proposed approach has the best performance with an F1-Score of 0.87 using the XGBoost algorithm. The other performance metrics of XGBoost have higher values than other models. The Accuracy of 0.91, Recall of 0.91, and Precision of 0.85 reveal the best performance of XGBoost among other models used in this research. It has also been observed that the proposed approach works well and can be used to help developers prioritize reviews for response.

#### 3.2.1 Case Study Evaluations

In this section, the best-trained model is evaluated on 8 popular applications. Firstly, 8 popular applications from different categories are selected and the selected model is tested on 1000 to 1500 reviews from each of the applications. After running the best model, the mean of F1 Score for all applications was 0.87 which is acceptable performance of the selected model. Also, according to the result, it can be concluded that the presented model is applicable for different types of applications from different categories.

### 3.3 Threats to Validity

In this section, the threats to the validity of our results are discussed. These threats are categorized into *internal*, *external*, *construct*, and *reliability* threats.

- **Internal Validity**: The accuracy and performance of the proposed approach are influenced by the features considered (such as the number of sentences, and the devices on which reviews are recorded) and the tools and libraries used. To mitigate these threats, the study uses well-known and accurate tools selected carefully.

- **External Validity**: The threat here is the potential inadequacy of collected data. This is addressed by using a crawler to extract reviews from many programs and merging this data with existing data from another source [7].

- **Construct Validity**: There is a concern about whether the textual and semantic features considered are relevant to the output. This is tackled by calculating the correlation between features and the output variable using Pearson's correlation coefficient, demonstrating that most selected features positively impact the prioritization of reviews.

- **Reliability**: This examines the consistency of results with similar inputs. The study ensures reproducibility through steps like preprocessing, feature extraction, and model training. Despite the probabilistic nature of machine learning, the study selects a diverse range of reviews to achieve a comprehensive understanding and mitigate uncertainty. The source code and dataset of the research are available at https://github.com/ISE-Research/App-Reviews-Prioritization.

## 4 Discussion

The evaluation results show that the XGBoost model has the best performance for prioritizing reviews with an F1-Score of 0.77 and 0.87 from Approach #1 and Approach #2. Therefore, developers can use the models presented in this study to prioritize user reviews and spend less time and effort responding to them.

In the analysis of our dataset, which consists of multiple attributes including 'rate', 'review length', 'readability score', 'review subjectivity', and several others, the XGBoost algorithm demonstrated superior performance compared to SVM, Decision Tree, and Random Forest algorithms. This can be attributed to XGBoost's ability to handle complex data interactions and its efficient implementation of gradient boosting, which allows it to capture intricate patterns and relationships within the data. Unlike SVM, which can struggle with large datasets and non-linear separations, XGBoost effectively scales with the dataset size and complexity. Similarly, while Decision Trees can easily overfit and Random Forests, though more robust, may not achieve the same level of accuracy and fine-tuning, XGBoost's advanced regularization techniques and hyperparameter optimization make it more adept at producing high-precision, high-recall, and high F1-Score metrics. This results in a more reliable and accurate predictive model for our multi-attribute dataset, highlighting XGBoost's efficacy and robustness in handling diverse and intricate data structures.

Nevertheless, like any other method, this work also exhibits both strengths and weaknesses. In terms of strengths, it introduces a novel categorization of user reviews based on two distinct approaches, incorporating a comprehensive array of textual and semantic features. It also identifies a substantial number of features crucial for prioritizing reviews for response, offering a fresh criterion for evaluating developers' perspectives. The study's validity is reinforced by its analysis of reviews spanning various application categories, and it carefully evaluates four machine learning algorithms to determine the most effective one for review prioritization. However, there are also weaknesses to consider. For instance, we exclude reviews updated after receiving a response due to limitations in accessing the initial posted date on Google Play. If we have a larger dataset, we can use deep learning techniques to achieve better results. Additionally, we can enhance the dataset with more features to obtain more precise outcomes.

## 5 Related Work

The studies conducted on the reviews are classified into several important areas: examining the relationship between users and developers, responding to user reviews, user review features and developer responses, automated response generation, user review classification, user review feature extraction, user sentiment analysis, user review analysis, and other divisions. Our main focus is on studies that examine user reviews and developer responses.

**Importance of Responding to User Reviews:** The study [17] concludes that as the number of downloads of an application increases, the number of responses developers provide to user reviews decreases, and conversely, as the number of downloads decreases, the number of responses to user reviews increases. It has also been noticed that users who initially rated an app with one star upgraded their rating to five stars after receiving a response.

**Automatic Response Generation:** The works in [25], [7], [9] examine various techniques used for responding to user reviews in apps, including linguistic patterns and key-word similarities. These studies are based on two main approaches. In the first approach, a tool is provided to developers to help them respond to user reviews on Google Play. This tool is the first automatic tool that does not rely on pre-prepared or rule-based responses; instead, it uses neural networks to generate responses. The second approach focuses on automatic response generation using neural machine translation networks. The article [11] presents ChatReview, a ChatGPT-enabled NLP framework designed to analyze domain-specific user reviews, offering personalized search results and addressing challenges such as bias and privacy.

**Feature Extraction from User Reviews:** [13], [8], and [27] focus on the automatic extraction of features from user reviews on internet sites. Finally, [13] deals with the extraction and matching of features mentioned in app descriptions and user reviews. The approach presented in [8] investigates the factors that determine the perceived helpfulness of user reviews. Specifically, it explores the correlation between user innovativeness and the content of reviews for innovative products, utilizing data gathered from a questionnaire survey. [27] aims to extract feature-describing phrases from app descriptions, align each app feature with its corresponding user reviews, and construct a regression model to determine which features exhibit significant associations with app ratings. Finally, [1] proposes an approach that assists in augmenting labeled datasets of app reviews by utilizing information extracted from GitHub issues that contain valuable information about user requirements.

**Sentiment Analysis of User Reviews:** In [15], a solution for employing sentiment analysis tools in software engineering datasets is presented. This work examines sentiment analysis in Stack Overflow discussions, JIRA issue reviews, and Google Play reviews. [22] employs a pre-trained RoBERTa language model for sentiment analysis and calculates cosine similarity for content-based recommendations. In [5], the impact of deep learning on sentiment analysis of Chinese reviews is examined.

**Users' Review Analysis:** This section presents the work related to the analysis of user reviews on internet sites. The work in [21] deals with automatic review analysis using clustering, TF-IDF, vector space model. [6] focuses on automatic insight extraction from user reviews over some time and prioritizing them using Gaussian distribution, linear regression, correlation coefficient, moving average, and Pearson correlation. [26] presents a tool for user review analysis using Word2Vec, vector space model, TF-IDF, and moving average. [14] presents MApp-IDEA tool to identify and categorize emerging concerns from user reviews, organizing them into a risk matrix with prioritization levels, and tracking their evolution over time. Finally, [18] presents UX-MAPPER, an approach to analyzing app store reviews and helping practitioners in identifying key elements influencing user experience.

Previous studies have used various machine learning methods to classify features from user reviews effectively. However, they have not explored using machine learning to prioritize reviews for responses. This research addresses this gap by collecting extensive datasets, identifying effective textual and semantic features for responding to reviews, and utilizing machine learning methods for prioritization.

# 6 Conclusions and Future Work

In recent years, the number of mobile applications in mobile app stores has been increasing, and every day a large number of users post their reviews and rate applications. Therefore, these users expect to receive a response from developers. Sometimes users increase their ratings after receiving a response from developers. As a result, to increase the app rating, it is necessary to respond to user reviews. However, responding to the large volume of reviews is one of the challenges facing developers, and they are not able to respond to all the reviews they receive. Therefore, it is necessary to prioritize user reviews to determine whether a developer should respond to a review or not. If so, it is important to know the response urgency associated with each review. In this paper, we proposed an approach based on natural language processing and machine learning techniques to prioritize users' reviews to respond. According to our evaluations, we observed that the XGBoost yields the most favourable outcomes with the best F1-Score with the value of 0.77 for Approach #1 and 0.87 for Approach #2 among other trained models for prioritizing user reviews regarding whether a response is necessary. Additionally, the XGBoost model demonstrates superior performance in determining the priority level of user reviews for response.

In the future, we plan to delve deeper into analyzing more textual and semantic features, like the number of sentences in a review, mentions of specific topics, and the types of devices used for leaving reviews. It would also be beneficial to test out different machine-learning algorithms to see how they compare in terms of performance. Additionally, it is important to see how well our approach works specifically within the Apple Store. Another useful step would be to categorize and group together reviews that share similar content, making it easier for developers to respond more effectively. Lastly, with the increased popularity of Large Language Models (LLMs) and Generative AI, we plan to work on generating automated responses to reviews to streamline the process even further.

*Engineering and Methodology*, July 2024.